\documentclass[prx,twocolumn,longbibliography,superscriptaddress,preprintnumbers]{revtex4-2}

\usepackage{graphicx}
\usepackage{dcolumn}
\usepackage{bm}
\usepackage{lineno}
\usepackage{ulem}
\usepackage{soul}
\usepackage{color}
\usepackage{array}
\usepackage{booktabs}
\usepackage{multirow, makecell}
\usepackage{physics}
\usepackage{lineno}
\usepackage[colorlinks=true, allcolors=blue]{hyperref}%
\usepackage{changes}
\usepackage{xcolor}
\usepackage{gensymb}

\newcommand{\RO}{RuO$_2$}

\begin{document}

\title{Correlation-driven tunability of altermagnetism in \RO
}

\author{Ina Park}
\affiliation{Center for Computational Quantum Physics (CCQ), Flatiron Institute, 162 5th Ave, New York, NY 10010, USA}

\author{Dongwook Kim}
\affiliation{Institute of Solid State Physics, TU Wien, 1040 Vienna, Austria}

\author{Inho Lee}
\affiliation{Theoretical Division, Los Alamos National Laboratory, Los Alamos, New Mexico 87545, USA}

\author{Jisook Hong}
\affiliation{POSCO Holdings, Pohang 37859, Republic of Korea
}

\author{Beomjoon Goh}
\email[]{beomjoon.goh@snu.ac.kr}
\affiliation{Institute for Data Innovation in Science, Seoul National University, Seoul 08826, Korea
}

\author{Bo Gyu Jang}
\email[]{bgjang@khu.ac.kr}
\affiliation{Department of Materials Science and Engineering, Kyung Hee University, Yongin 17104, Republic of Korea
}

\date{\today}

\begin{abstract}
\RO\ has been regarded as a prototypical candidate for metallic altermagnet, offering a potential platform for high-speed and high-efficiency spintronics. However, the magnetic ground state of \RO\ remains a topic of active debate due to conflicting experimental reports. In this work, we investigate the effect of electron correlations in \RO\ using density functional theory combined with dynamical mean-field theory (DFT+DMFT). In contrast to previous DFT-based studies, DFT+DMFT captures essential dynamical correlation effects, yielding spectral functions and optical conductivities in excellent quantitative agreement with experiments, and further reveals that \RO\ resides in the close vicinity of both the paramagnetic--altermagnetic phase boundary and the itinerant--localized crossover, rendering the magnetic ground state highly susceptible to external perturbations. 
Indeed, even a minimal compressive strain of $\sim$0.5\% is sufficient to drive the system into an altermagnetic phase. These findings elucidate the origin of the conflicting experimental observations and reveal that dynamical correlation effects are the key driving force behind the highly tunable magnetic ground state of \RO.
\end{abstract}

\maketitle


\section{Introduction}
Altermagnetism, recently established as a distinct class of collinear magnetism~\cite{vsmejkal2022beyond,vsmejkal2022emerging, jungwirth2026symmetry}, has emerged as a promising platform for spintronics applications~\cite{vsmejkal2022emerging, vsmejkal2022giant,Jungwirth2025spintronics, song2025altermagnets}. It uniquely integrates advantageous features of both antiferromagnets and ferromagnets, namely, THz-range dynamics and zero stray fields, together with non-relativistic spin-splitting and strong spin-dependent responses. This combination makes altermagnets particularly appealing for future spintronic devices based on spin-polarized currents, spin-transfer torque, and giant (GMR) and tunneling magnetoresistance (TMR), with potential applications for non-volatile magnetic memory and energy-efficient spin transistors and filters.

Ruthenium dioxide (\RO)~\cite{vsmejkal2020crystal, gonzalez2021efficient, bose2022tilted, bai2022observation, karube2022observation, fedchenko2024observation} is one of the most extensively studied altermagnetic (AM) material candidates. (Crystal structure shown in Fig.~\ref{comparison}(a).) Had long been identified as a Pauli paramagnet, it was later reinterpreted as an itinerant antiferromagnet with small but finite magnetic moment of 0.05 ($\mu_{\rm B}$/Ru) and  high N\'eel temperature of $T_N\sim$ 300 K from neutron scattering and X-ray studies~\cite{berlijn2017itinerant, zhu2019anomalous}. More recently, with the emergence of the altermagnetism field, theoretical studies have proposed \RO\ as hosting a spin-split electronic structure with $d$-wave spin texture, i.e., altermagnet~\cite{vsmejkal2020crystal,vsmejkal2022emerging}. Density functional theory (DFT) calculations predicted a remarkably large non-relativistic spin splitting size of $\sim$1 eV near the Fermi level~\cite{vsmejkal2022emerging} and potential GMR and TMR~\cite{vsmejkal2022giant, vsmejkal2020crystal},  placing \RO\ at the forefront of metallic altermagnet candidates.

Early experiments such as angle-resolved photoemission spectroscopy (ARPES), magnetic circular dichroism (MCD)~\cite{fedchenko2024observation}, anomalous Hall responses~\cite{feng2022anomalous, tschirner2023saturation}, and spin current measurement~\cite{bose2022tilted, bai2022observation, karube2022observation} supported the proposed AM phase in \RO. However, more recent experiments~\cite{hiraishi2024nonmagnetic, kessler2024absence, liu2024absence, wenzel2025fermi, wu2025fermi, kiefer2025crystal, peng2025universal} have reported results that are difficult to reconcile with a robust AM ground state in bulk \RO. For example, spin-resolved ARPES (SARPES)~\cite{liu2024absence} and optical conductivity measurements~\cite{wenzel2025fermi} together with comparisons to DFT(+$U$) have not provided clear evidence for spin-split electronic structure. 

These contradictory observations propose a shift in perspective: rather than focusing on whether \RO\ is intrinsically AM or paramagnetic (PM), it would be more meaningful to explore under what conditions an AM phase can be stabilized. Indeed, various extrinsic factors such as epitaxial strain, dimensionality, vacancy-induced doping, and defects have been discussed as possible factors.~\cite{smolyanyuk2024fragility, forte2025strain, kiefer2025crystal, li2026exploration, song2025altermagnets} Notably, while many bulk measurements challenge the existence of AM phase, some of recent thin-film and ultra thin-film studies continue to report observations compatible with existing AM electronic structure~\cite{liu2023inverse,jung2025reversible, sun2025chiral, li2025fully}. Recent theoretical studies based on DFT+$U$ also suggested possible AM ground state under strain or doping~\cite{smolyanyuk2024fragility, forte2025strain}.

On the other hand, however, most theoretical comparisons relied on DFT(+$U$) which treats electron correlations at a static mean-field level. While widely used and useful, they inherently lack the dynamical correlation effects that can be important for quantitative comparisons with spectroscopic and transport probes like ARPES or optical conductivity. In particular, the effects of dynamic electronic correlations can be important in 4$d$ transition-metal oxides, including \RO. They are usually located in an intermediate regime between localized and itinerant regimes, where electrons are more itinerant than in typical 3$d$ oxides, yet remain significantly affected by local Coulomb interactions $U$ and Hund’s coupling $J$ ~\cite{mravlje2011coherence,georges2013strong, wadati2014photoemission, kim2015nature, Deng2016, Zhang2016_sr2ruo4, Kim2018_sr2ruo4, Tamai2019, Kim2022}.

In this context, \RO\ can be considered as a moderately correlated metal, which shows clear Fermi liquid behavior~\cite{ling2026t} with effective masses ranging from 2 to 5 measured from the de Haas van Alphen (dHvA) experiment for different orbits~\cite{wu2025fermi} and an effective mass of $\sim$ 1.5 estimated from optical conductivity~\cite{wenzel2025fermi} and heat capacity measurement~\cite{ling2026t}. In such cases, a method that can account for both itinerant properties and local correlation effects, such as DFT+DMFT, is required. DFT+DMFT incorporates a frequency-dependent self-energy to the electronic structure which can therefore describe low-energy quasiparticle behavior and spectral weight redistribution as well as capturing local spin fluctuations that can be important in multi-orbital systems such as \RO.

In this work, we revisit the electronic structure and magnetic ground state of bulk and strained \RO\ using DFT+DMFT. We clearly demonstrate that dynamic correlations are essential to reliably compare to the experimental results such as ARPES and optical conductivity. The results further reveal that the bulk \RO\ is not only near the PM--AM boundary but also near the itinerant-to-localized boundary. Crucially, the dynamical correlation-induced band renormalization shifts the flat band closer to the Fermi level, enhancing the density of states and driving a magnetic instability, such that even a minimal compressive strain readily stabilizes the AM phase. This intrinsic proximity to phase boundaries elucidates why the magnetic ground state of \RO\ is highly tunable under external perturbations.


\section{I. Correlation effects in bulk {\protect\RO}}

\begin{figure*}
\includegraphics[width=\textwidth]{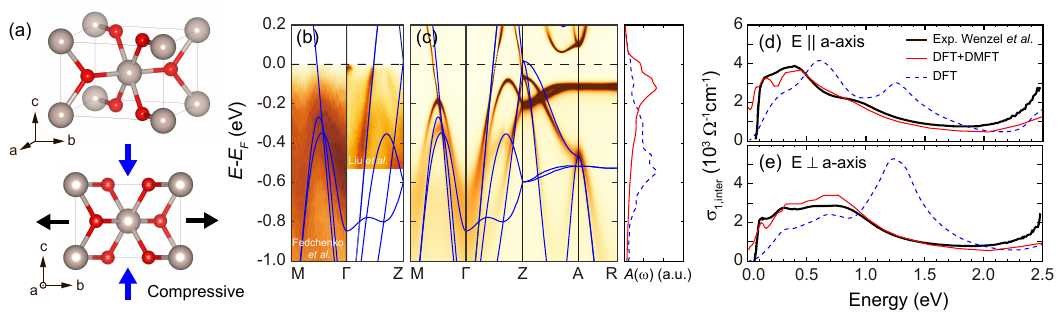}
\caption{(a) Crystal structure of rutile \RO. (Top) Pristine crystal structure without strain. (Bottom) Structure with compressive strain along $c$ direction. (b) Experimental ARPES intensity map~\cite{fedchenko2024observation, liu2024absence} overlaid with NM DFT band structure (blue lines) (c) PM DFT+DMFT spectral function $A(\mathbf{k},\omega)$ with DFT bands (blue lines). (Right panel) the Ru $d_{x^2-y^2}$-projected momentum-integrated spectral function $A(\omega)$. (d-e) Interband optical conductivity $\sigma_{1,inter}$($\omega$) for (d) $E\parallel a$ and (e) $E \perp a$ polarization directions, comparing experiment~\cite{wenzel2025fermi}, DFT+DMFT (red), and DFT (blue dashed). }
\label{comparison}
\end{figure*}



Figure~\ref{comparison}(b) shows the experimental ARPES intensity map of \RO\ along the M--$\Gamma$--Z high-symmetry path, reproduced from Liu $et\ al$.~\cite{liu2024absence} and Fedchenko $et\ al$.~\cite{fedchenko2024observation}, overlaid with the non-magnetic (NM) DFT band structure (blue lines). (See Methods for calculation details.) While the DFT bands capture the overall dispersion observed in the experiments, there are notable discrepancies between them. First, the experimental spectral weight is shifted to lower binding energies relative to the DFT prediction, indicating the band renormalization. For instance, the band appearing near -0.8~eV at the $\Gamma$ point in DFT is observed at approximately -0.6~eV in the ARPES intensity. In addition, the band crossing the Fermi level ($E_F$) along the $\Gamma$--Z path exhibits a smaller slope compared to the DFT prediction. Second, the experimental spectral features are significantly broadened and incoherent. The band renormalization and reduced quasiparticle lifetime clearly indicate the electron correlation in this system.

Figure~\ref{comparison}(c) presents the PM DFT+DMFT spectral function along the extended high-symmetry path. The DFT+DMFT calculations were performed at $T\sim$58 K with $(U,J) = (3.5, 0.6)$~eV. The interaction parameters, Coulomb $U$ and Hund $J$, used in calculations were carefully determined based on previous DFT+DMFT studies on ruthenate materials and recent estimates obtained from constrained DFT and DMFT calculations~\cite{kim2015nature, Deng2016, Zhang2016_sr2ruo4, Kim2018_sr2ruo4, Tamai2019, Kim2022, hauleconstrained}. Notably, the electron-pocket-like band around $\Gamma$ point becomes strongly incoherent with substantial band renormalization, leading to good agreement between ARPES intensity and DFT+DMFT spectral function. 
The effective mass $m^*/m$ obtained from DFT+DMFT calculation is $\sim1.4$, which agrees well with that obtained from experiments~\cite{wenzel2025fermi, ling2026t}
($m^*/m = 1 - \partial\,\mathrm{Im}\,\Sigma(i\omega)/\partial\omega|_{\omega\rightarrow0^+}$, where $\mathrm{Im}\,\Sigma(i\omega)$ is the imaginary part of the DMFT self-energy on the Matsubara frequency).
Another interesting feature is the band-dependent renormalization. While the hole-pocket-like bands between M and $\Gamma$ point and at the A point exhibit relatively weaker renormalization, the flat bands along the Z--A--R path are strongly renormalized. 
The right panel of Fig.~\ref{comparison}(c) show the partial density of states (PDOS) of flat band orbital, Ru $d_{x^2-y^2}$ (See the Appendix~\ref{sec:fatband} for orbital notation), obtained from DFT (blue dashed) and DFT+DMFT (red), respectively. One can clearly observe that band renormalization shifts the flat band closer to $E_F$, thereby enhancing the density of states and bringing the system closer to a magnetic instability.

The effect of electron correlations in \RO\ is more pronounced in the optical conductivity. Figure~\ref{comparison}(d) and (e) show the interband optical conductivity for $E \parallel a$ and $E \perp a$, respectively, with the experimental data reproduced from Ref~\cite{wenzel2025fermi}. The optical conductivity obtained from the NM DFT calculation (blue dashed line) exhibits significant deviation from the experiment, whereas PM DFT+DMFT result (red line) is in excellent agreement with the experimental data for both polarization directions. For $E \parallel a$ direction, notable peaks at 0.7 eV in the DFT calculation are shifted to 0.4 eV in both the experimental data and the DFT+DMFT result. In addition, the sharp peak structures in the DFT results are smoothed out in the experimental data and the DFT+DMFT result, most notably for the peak at 1.3 eV for the $E \perp a$ direction.
This prominent peak at $\sim$1.3~eV in the DFT optical conductivity originates from interband transitions along the $Z-A$ path, where bands are separated by $\sim$1.3--1.4~eV (See Appendix Fig.~\ref{fig:afmoptics}(a) for the band structure in the extended energy window). In the DFT+DMFT spectral function, however, the high-energy states involved in these transitions become strongly incoherent, leading to a significant suppression of the corresponding spectral weight and hence the optical transition. The peak shift (band renormalization) and the peak suppression (finite quasiparticle lifetime) clearly demonstrate the importance of dynamical correlation effects in \RO.

It is worth noting that a previous study reported a good agreement between the NM DFT optical conductivity and the experiment, while the AM DFT+$U$ ($U$=2 eV) result was found to deviate from the experimental data~\cite{wenzel2025fermi}. (We adopt the prefix `AM' to maintain terminological consistency with the material's altermagnetic phase, e.g. AM DFT+$U$. The calculations are set with antiparallel spin configuration on spin sublattices, identical to a standard AFM setup.) However, this comparison was made after rescaling the frequency axis of the NM DFT result by a factor of 1.3, mimicking the effect of quasiparticle renormalization, and also applying a frequency-dependent broadening to account for the finite quasiparticle lifetime~\cite{wenzel2025fermi}, both of which are dynamical correlation effects that are absent in static DFT calculations. Although such an energy rescaling is not uncommon in the DFT study of correlated metallic systems, it is difficult to draw a definitive conclusion on which ground state, NM or AM, better describes the experimental optical conductivity based on this approach. Indeed, the AM DFT+DMFT optical conductivity exhibits no significant deviation from the PM result (See Appendix Fig.~\ref{fig:afmoptics}(b)). It also yields an ordered moment of  $\sim0.1~\mu_B$/Ru, consistent with the small magnetic moment observed experimentally~\cite{berlijn2017itinerant}.
These results imply that determining the ground state solely from a comparison between experiment and DFT (or DFT+$U$) without dynamical correlation effects is not conclusive (The AM solution is marginally less stable than the PM solution in DFT+DMFT calculations as discussed in the following section).


\begin{figure*}
\includegraphics[width=\textwidth]{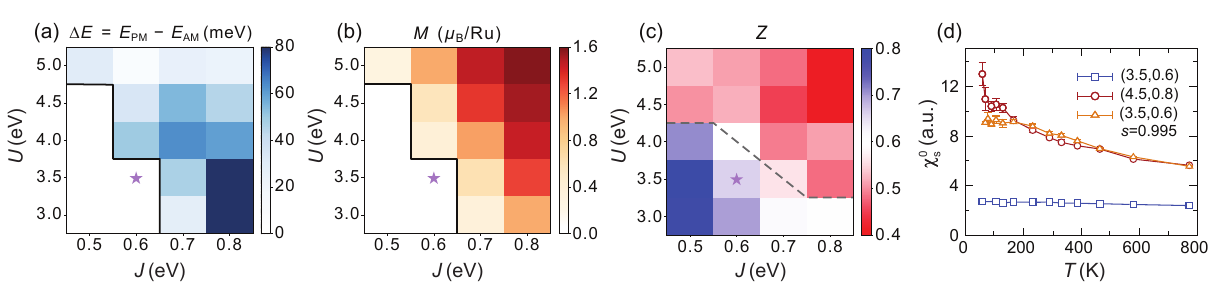}
\caption{DFT+DMFT $U-J$ phase diagram for (a) the energy difference between the PM and AM phase, $\Delta E = E_{\rm{PM}} - E_{\rm{AM}}$ (meV/f.u.), (b) ordered moment $M$ ($\mu_\mathrm{B}$/Ru). The black solid line indicates the phase boundary between PM and AM phase. (c) Quasiparticle weight $Z$ of Ru $d_{x^2-y^2}$ orbital. The itinerant to localized crossover is guided by the grey dashed line. (d) local spin susceptibility $\chi_S^0$ vs. $T$ for pristine (square) and 0.5\% strained ($s=0.995$) (triangle) \RO\ with chosen interaction parameters $(U,J) = (3.5, 0.6)$ (eV) and pristine \RO\ in localized limit with $(U,J) = (4.5, 0.8)$ (circle). Purple star marks the chosen interaction parameter used for the comparison between DFT+DMFT calculation and experiments.} 
\label{phasediagram}
\end{figure*}

\section{II. {\protect\RO} in the vicinity of Phase boundary}

To further investigate the effect of electron correlations in \RO, we performed PM and AM DFT+DMFT calculations with various $(U,J)$ parameters. To exclude the structural effect, all calculations were performed based on the fixed crystal structure. (See Methods.) 
Figure~\ref{phasediagram}(a) shows the $U-J$ phase diagram of the energy difference between PM and AM states. 
Bluish colors in the figure indicate that the AM solution is energetically more stable than the PM solution, with darker shades corresponding to a larger energy difference. For small $J$ of 0.5 or 0.6~eV, the AM solution is not stabilized unless $U$ is sufficiently large, whereas for large $J$, the AM solution is readily stabilized even with small $U$ values. The black solid line indicates the phase boundary line between NM and AFM states. It should be noted that the $U-J$ parameter set used in Fig.~\ref{comparison}, which well reproduces the experimental results, is located at the phase boundary as indicated by a star, where the PM solution is marginally more stable than the AM solution.
 This suggests that the ground state of bulk \RO\ is not AM but can be highly sensitive to external perturbations.

This energy phase diagram also reveals essential features that are inaccessible within static DFT+$U$ calculations. In DFT+$U$ calculations, the AM solution becomes increasingly stabilized as $U$ increases~\cite{smolyanyuk2024fragility, forte2025strain, meinert2025meta, hou2026nonmagnetic}. However, DFT+DMFT calculations do not exhibit such monotonic behavior in the energy difference, even though the ordered moment $M$ in Fig.~\ref{phasediagram}(b) increases monotonically with both $U$ and $J$, like DFT+$U$ calculations~\cite{forte2025strain, meinert2025meta, hou2026nonmagnetic}. For instance, the energy difference becomes smaller at large $U$ values, in contrast to DFT+$U$ calculations. This non-monotonic behavior is most clearly seen for $J=0.7$~eV, where the AM solution becomes increasingly stabilized with increasing $U$ at first, but the energy difference starts to decrease above $U=4$~eV despite the AM solution remaining stable. For $J=0.8$~eV, the AM solution is strongly stabilized already at $U=3$~eV, and the energy difference continuously decreases as $U$ increases.

To understand this behavior, it is helpful to examine the quasiparticle weight ($Z = (m^*/m)^{-1}$) phase diagram obtained from PM DFT+DMFT calculations, shown in Fig.~\ref{phasediagram}(c). 
In the figure, the $Z$ of $d_{x^2-y^2}$ orbital is shown, which dominantly contributes to the flat band near $E_F$ and is sensitive to changes in $U$ or $J$. The first thing to note is that $Z$ is rapidly suppressed with increasing $U$ at small $U$ values, but the further suppression becomes marginal at larger $U$ values. For instance, at $J=0.6$ eV, $Z$ is suppressed from 0.7 to 0.5 as $U$ increases from 3 to 4.5~eV and is not further suppressed at $U=5$~eV. The same trend is also observed for $J=0.7$ eV, where $Z$ decreases from 0.62 to 0.5 as $U$ increases from 3 to 4~eV and shows no significant further suppression above $U=4$~eV. (See Appendix Fig.~\ref{fig:Z} for raw data.) This is the hallmark of Hund's metal behavior in non-singly-occupied and non-half-filled systems, where the low- and high-energy effects act in opposite directions, giving rise to the so-called Janus-faced character of Hund's correlation~\cite{de2011janus, georges2013strong, stadler2015dynamical, georges2024hund}. 
Specifically, at low energies, $J$ suppresses the Kondo screening scale, leading to incoherent and renormalized quasiparticles. At high energies, on the other hand, $J$ reduces the charge transfer energy, making the system less correlated in terms of charge fluctuations~\cite{georges2013strong, stadler2015dynamical, georges2024hund}. This, in turn, gives a rapid decrease of $Z$ at smaller $U$ followed by a slow decrease of $Z$ at larger $U$, making the system correlated but keeping it away from the Mott insulating phase.

\begin{figure*}
\includegraphics[width=\textwidth]{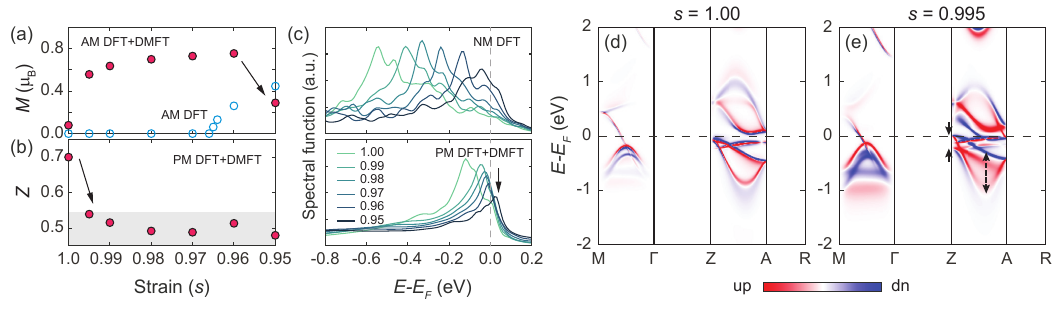}
\caption{Strain-dependent properties of \RO\ obtained from DFT (blue) and DFT+DMFT (red). (a) Ordered magnetic moment of Ru atom as a function of strain. (b) Quasiparticle weight $Z$ of Ru $d_{x^2-y^2}$ orbital as a function of strain. (c) Partial density of states (PDOS) of the flat band orbital ($d_{x^2-y^2}$) as a function of strain, with the upper and lower panels showing the NM DFT and PM DFT+DMFT results, respectively. Darker colors correspond to larger compressive strain. (d,e) Spin-resolved spectral function $A(\mathbf{k},\omega)$ obtained from AM DFT+DMFT calculations for (d) zero strain $s=1.0$ and (e) finite compressive strain $s=0.995$. Red and blue marks spin-up (up) and spin-down (dn) components, respectively. Dashed and solid arrow marks the AM spin splitting and exchange splitting size, respectively, for one of representative bands near the Fermi level. (Refer main text for details.)}
\label{strain}
\end{figure*}

Together with the $Z$~phase diagram, the non-monotonic behavior in the energy phase diagram (Fig.~\ref{phasediagram}(a)) can be understood in terms of an itinerant-to-localized transition. 
At small $U$ values (itinerant regime), the local fluctuating moment is not sufficiently large. As $U$ increases, the moment grows, accompanied by a gradual decrease in $Z$ and a progressive stabilization of the AM solution. However, once $Z$ drops below a critical value of $\sim$0.5, driving the system into the localized regime, the exchange energy $J_{\mathrm{ex}}$
begins to decrease with further increasing $U$ ($J_{\mathrm{ex}} \sim t^2/U$), which in turn reduces the energy difference between the AM and PM solutions. 
Therefore, pristine \RO\ is located at the itinerant--localized boundary, guided by the grey dashed line in Fig.~\ref{phasediagram}(c). 
As the system crosses into the localized regime, the AM solution becomes stabilized, which appears to be closely related to the position of the flat band discussed later. 

The itinerant-to-localized transition is further confirmed in the temperature ($T$)-dependent local spin susceptibility $\chi_S^0$ obtained from PM DFT+DMFT calculations as shown in Fig.~\ref{phasediagram}(d). For interaction parameter $(U,J) = (3.5, 0.6)$~eV (squares), which well reproduces the experimental spectral functions and optical conductivity, $\chi_S^0$ is nearly $T$-independent over the entire temperature range up to 800~K, indicating the itinerant nature. In contrast, $\chi_S^0$ from $(U,J) = (4.5, 0.8)$~eV calculations exhibits Curie--Weiss behavior (circles), indicating the formation of local magnetic moments. Taken together, the energy phase diagram, the $Z$ phase diagram, and $T$-dependent $\chi_S^0$ clearly indicate that \RO\ is located in the close vicinity of not only the PM--AM boundary but also the itinerant--localized boundary. These results provide a natural explanation for the conflicting experimental observations on the magnetic ground state of \RO\ and indicate that dynamical correlation effects are a key driving force behind the highly tunable magnetic ground state.

\section{III. Correlation-Driven Sensitivity of the AM Phase to Strain}

In light of these findings, 
we further examine how sensitively the correlation strength and related magnetic ground state of \RO\ can be tuned by strain, using DFT+DMFT with the same interaction parameter set as in Fig.~\ref{comparison}. Figure~\ref{strain}(a) shows the ordered moment $M$ obtained from AM DFT and DFT+DMFT calculations as a function of strain along the $c$ axis.
In this paper, the compression ratio of the $c$ lattice constant is denoted by $s$, where $s=0.99$ corresponds to 1\% reduction in $c$. The in-plane lattice constant and internal atomic positions are fully relaxed accordingly for each strained case. (See Methods.)
At zero strain ($s=1.0$), since the energy difference between the AM and PM solutions is very small in the DFT+DMFT calculations, the ordered moment obtained from the AM solution ($\sim0.1~\mu_{\rm{B}}$/Ru) is depicted to provide a clearer picture of how sensitive the moment evolves with strain. In the DFT calculation, a compressive strain below $s=0.97$ is required for the AM moment to emerge. In contrast, even a small strain immediately induces a sharp onset of the ordered moment in the DFT+DMFT calculation. 

The spin-resolved spectral functions for $s=1.0$ and $s=0.995$ cases are shown in Figs.~\ref{strain}(d) and \ref{strain}(e). In these figures, spin-up and spin-down spectral functions are exactly the same along the symmetry-protected nodal lines, $\Gamma$--Z and A--R. 
Compared to the zero-strain case, 0.5\% of compressive strain ($s=0.995$) already exhibits a much sharper contrast between the spin-up and spin-down spectral functions, together with larger AM spin-splitting. 
It is noteworthy that in addition to AM spin splitting due to anisotropic crystal potential (dashed arrow), the exchange splitting (solid arrow), isotropic spin splitting within the same spin sublattices~\cite{vsmejkal2022emerging}, significantly increases already with a small amount of strain. In the case of 0.5\% compressive strained case, the AM spin-splitting and exchange splitting are obtained as $\sim$0.8 eV and $\sim$0.3 eV, respectively.

Interestingly, the abrupt increase in the ordered moment is concurrently accompanied by an abrupt decrease in the quasiparticle weight $Z$ in PM DFT+DMFT calculations. Applying a strain of merely $s=0.995$ causes $Z$ to drop sharply from $\sim$0.7 to $\sim$0.5, indicated by the black arrow in Fig.~\ref{strain}(b).
This abrupt decrease in $Z$ is reflected in the PDOS of the flat band $d_{x^2-y^2}$ orbital shown in Fig.~\ref{strain}(c), where the PM DFT+DMFT result shows an abrupt shift of the flat band peak toward $E_F$ at $s=0.995$ compared to the gradual evolution seen in NM DFT. Since the flat band is already located near $E_F$ in DFT+DMFT calculations due to band renormalization, the additional reduction in $Z$ at $s=0.995$ further accelerates this shift toward $E_F$. This enhances the density of states near $E_F$ and triggers the magnetic instability. In Fig.~\ref{phasediagram}(d), the local spin susceptibility $\chi_S^0$ of the $s=0.995$ case (triangles) clearly exhibits temperature dependence similar to that observed in the localized regime of the unstrained case (circles), confirming that the system has crossed the itinerant--localized boundary. Notably, $\chi_S^0$ of the $s=0.995$ case shows a saturation below $\sim$150~K, suggesting that the local moment is screened at low temperatures as the system resides just across this boundary.

It is worth noting that the ordered moment emerges when the flat band position moves above $-0.15$~eV in both the NM DFT and PM DFT+DMFT calculations, which occurs at $s=0.965$ in DFT but just above zero strain in DFT+DMFT, reflecting the much smaller critical strain required due to correlation-induced band renormalization. Another interesting feature is that the ordered moment is sharply suppressed as soon as the flat band is pushed above $E_F$ at $s=0.95$ (See black arrow in Fig.~\ref{strain}(a) and (c)), with the same trend observed in DFT but at a significantly larger compressive strain of $\sim$0.90. These results demonstrate that the position of the flat band is strongly coupled to the stabilization of the AM state, and that dynamical correlation-induced band renormalization renders the system highly susceptible to magnetic instability under external perturbations.

\section{Conclusion}
In summary, we have investigated the magnetic ground state and electronic structure of bulk and strained \RO\ using DFT+DMFT. Careful comparison with recent ARPES and optical conductivity data demonstrates that dynamical correlation effects, including band renormalization and finite quasiparticle lifetime, are essential for a quantitative description of \RO. While DFT+DMFT identifies the bulk ground state as paramagnetic, it reveals that \RO\ resides in the close vicinity of both the paramagnetic--altermagnetic phase boundary and the itinerant--localized crossover, which is only captured when dynamical correlations are properly taken into account. This proximity renders the magnetic ground state highly susceptible to external perturbations, such that even a minimal compressive strain of $\sim$0.5\% is sufficient to drive the system into an AM phase. These findings provide a natural and unified explanation for the conflicting experimental observations and demonstrate that dynamical electron correlations are responsible for the highly tunable magnetic ground state of \RO, providing a pathway toward realizing and controlling the AM phase for spintronics applications.
\vspace{2em}

\noindent
{\it Methods} -- 
For the pristine structure, the experimental lattice constants were used~\cite{kiefer2025crystal}, and only the internal atomic positions were relaxed using NM DFT calculations. The relaxed internal atomic positions are in good agreement with the experimental values. For the strained structures, the $c$-axis lattice constant was systematically reduced from the experimental value and fixed, while the in-plane lattice constant and internal atomic positions were fully relaxed accordingly.
All structure relaxations were performed using the Vienna \textit{ab\ initio} Simulation Package (\texttt{VASP}), which employs the projector augmented-wave (PAW) method~\cite{vasp1, vasp2}. 
The Perdew--Burke--Ernzerhof (PBE) generalized gradient approximation (GGA) was employed for the exchange-correlation functional~\cite{gga}. A plane-wave cutoff was set to 500 eV, and a $7\times7\times11$ $k$-point mesh was used.

Based on the relaxed structure, the electronic structure was calculated using the \texttt{WIEN2k}, which employs the full-potential augmented plane wave method~\cite{Blaha2020}. A 2000 $k$-point mesh was used for self-consistent calculation. On top of an effective one-electron Hamiltonian obtained from the \texttt{WIEN2k} calculation, charge self-consistent DFT+DMFT calculations were performed as implemented in the DFT + embedded DMFT functional (eDMFTF) code~\cite{haule2010dynamical, haule2015free}. For the DMFT calculations, a real harmonics basis for Ru 4$d$ orbitals was used with local axis rotation corresponding to the rutile structure. Further local basis transformation was also applied to describe the distorted octahedral ligand field environment generated by O atoms~\cite{occhialini2021local}. (Refer to Appendix for details and Fig.~\ref{fig:fatband} for the resulting orbital projected band structures.) For the interaction Hamiltonian, we used the density-density form of the Coulomb interaction with Slater parametrization, $F^0 \equiv U$, $F^2 \equiv \frac{112}{13}J$, and $F^4 \equiv \frac{70}{13}J$ with interaction parameters $U$  and $J$. Various $U$ and $J$ parameters were used for scanning the phase diagrams shown in Fig.~\ref{phasediagram}, and $U=3.5$ eV and $J=0.6$ eV were used for all the comparisons to experimental results shown in Fig.~\ref{comparison} and for the strained case shown in Fig.~\ref{strain}. These interaction parameters are carefully chosen as discussed in the main text, based on previous studies~\cite{kim2015nature, hauleconstrained}.
The hybridization window was set from -10 eV to 10 eV, and the impurity problem was solved by a continuous-time quantum Monte Carlo (CTQMC) impurity solver. 
\vspace{1em}

\noindent
{\it Acknowledgements} -- I.P. thank Se Young Park, Changhee Sohn, and Harrison Labollita for fruitful discussions. B.G. is supported by the Global-LAMP Program of the National Research Foundation of Korea (NRF) grant funded by the Ministry of Education (No. RS-2023-00301976). B.G.J. is supported by the National Research Foundation of Korea(NRF) grant funded by  the Korea government (MSIT) (RS-2026-25488595), the Center for Advanced Computation (CAC) at Korea Institute for Advanced Study (KIAS) and the National Supercomputing Center with supercomputing resources including technical support (KSC-2025-CRE-0367). Work at Los Alamos was carried out under the auspices of the U.S. Department of Energy (DOE) National Nuclear Security Administration (NNSA) under Contract No. 89233218CNA000001. I.L. was supported by the LANL LDRD Program, and in part by Center for Integrated nanotechnologies, a DOE BES user facility, in partnership with the LANL Institutional Computing Program for computational resources. The Flatiron Institute is a division of the Simons Foundation.
\vspace{1em}

\noindent
{\it Supporting Information} -- Supporting Information is available.
\vspace{1em}

\noindent
{\it Data Availability} -- Data are available from the authors upon reasonable request.

\bibliographystyle{apsrev4-2}
\bibliography{reference.bib}

\clearpage

\appendix

\setcounter{section}{0}         

\begin{figure*}[h!]
    \centering
    \includegraphics[width=\linewidth]{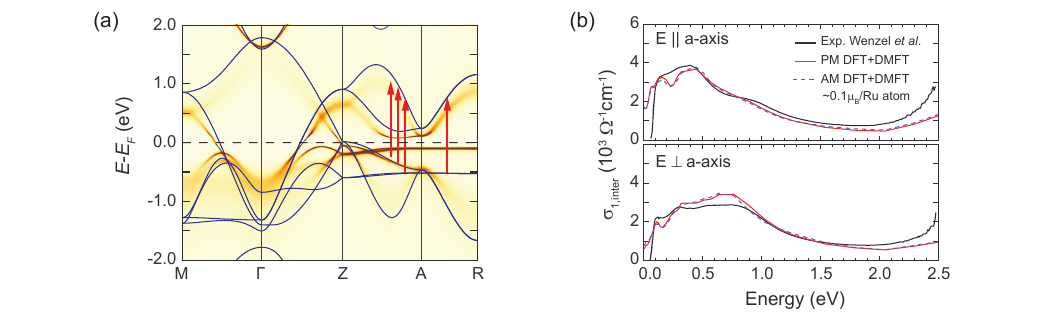}
    \caption{(a) DFT and DFT+DMFT spectral function in the extended energy
window. Red arrows indicate interband transitions with energy separation of 1.3~eV, corresponding to the prominent peak observed in the NM DFT optical conductivity (Fig.~\ref{comparison}(d) and (e)). (b) Interband optical conductivity from AM DFT+DMFT calculation (dashed) with ordered moment $M\sim0.1\mu_B/\rm{Ru}$ compared with PM DFT+DMFT (thin red) and experimental result (thick black)~\cite{wenzel2025fermi}. }
    \label{fig:afmoptics}
\end{figure*}

\section{DFT+DMFT Spectral Functions and Optical conductivity\label{sec:transition}}

Fig.~\ref{fig:afmoptics}(a) shows the PM DFT+DMFT spectral function and DFT band structure over an extended energy window. The red arrows, each spanning $\sim$1.3~eV, indicate interband transitions along the Z--A path that correspond to the prominent peak observed in the NM DFT optical conductivity. Nearly parallel bands along the Z--A direction, separated by $\sim$1.0--1.5~eV, collectively contribute to this peak, with the flat band also appearing to play a role. In the DFT+DMFT spectral function, the high-energy states involved in these transitions become strongly incoherent, suppressing the corresponding optical transition. Simultaneously, the flat band undergoes significant renormalization, such that the sharp peak at $\sim$1.3~eV is absent in the DFT+DMFT result and is replaced by a broad shoulder near $\sim$0.9~eV. Fig.~\ref{fig:afmoptics}(b) shows the interband optical conductivity for the PM and AM states calculated with $(U,J)$ = (3.5, 0.6)~eV. The AM state is only marginally less stable than the PM state, yielding a small ordered moment of $\sim0.1~\mu_B$/Ru. The optical conductivity of the AM and PM states are nearly indistinguishable. This demonstrates that a comparison between the experimental optical conductivity and DFT(+$U$) calculations alone is insufficient to draw a conclusive determination of the magnetic ground state.

\section{DFT+DMFT orbital-projected band structures\label{sec:fatband}}
\begin{figure*}[h!]
    \centering
    \includegraphics[width=0.99\linewidth]{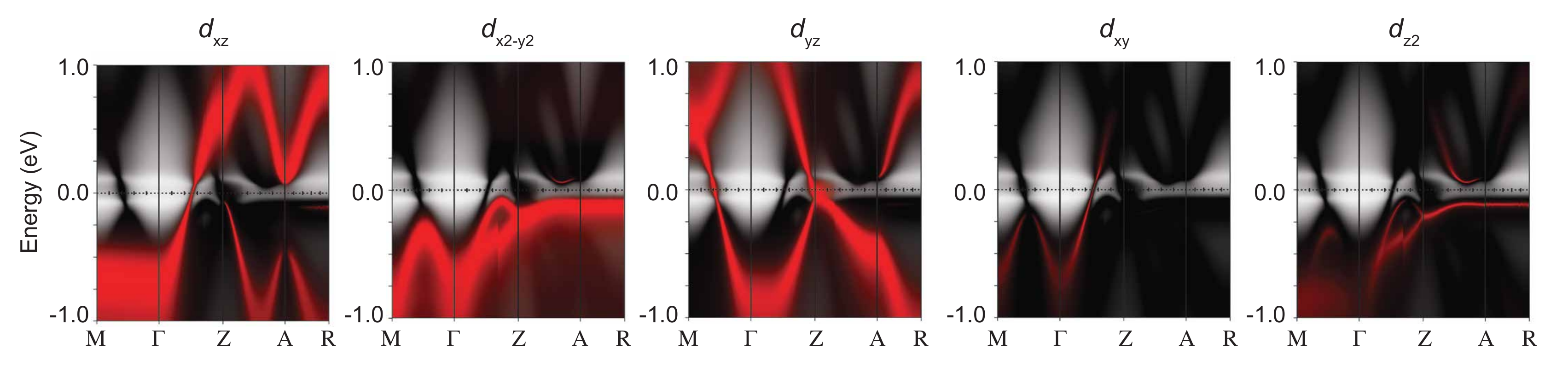}
    \caption{DFT+DMFT orbital-projected band structure of Ru1 atom in the case of pristine \RO. Refer the section note for orbital nomenclature.}
    \label{fig:fatband}
\end{figure*}

Fig.~\ref{fig:fatband} shows the Ru1 orbital-projected band structure, momentum-resolved spectral function $A(k,\omega)$, obtained from PM DFT+DMFT calculation for pristine \RO. In DFT+DMFT calculation, for Ru 4$d$ orbitals, the local axis rotation corresponding to rutile structure together with the orbital basis transformation corresponding to the nearly-octahedral ligand field environment to maximally diagonalize the impurity Green's function was applied. 

In rutile structure, the ligand field environment of Ru 4$d$ orbitals does not form a perfect octahedron, as in-plane O--Ru--O angles are 103$\degree$ and 77$\degree$, respectively. As a result, the distorted octahedral ligand field environment first introduces a large octahedral crystal field splitting, with $\Delta_{e_g-t_{2g}}\sim 2$ eV, then further symmetry lowering from $O_h$ to $D_{4h}$ local symmetry point group occurs, breaking the $t_{2g}$ orbital degeneracy with the corresponding crystal field splitting energy scale much larger than the spin-orbital coupling energy scale~\cite{occhialini2021local}.

Accordingly, the orbital names assigned in Fig.~\ref{fig:fatband} - $d_{xz}$, $d_{x^2-y^2}$, $d_{yz}$, $d_{xy}$, and $d_{z^2}$~\cite{occhialini2021local} - are linear combinations of the original real harmonics $d$ orbital basis functions. As shown in Fig.~\ref{fig:fatband}, the resulting basis transformation gives two less-filled $e_g$ orbitals - named as $d_{xy}$ and $d_{z^2}$ - and three more-filled non-degenerate $t_{2g}$ orbitals - named $d_{xz}$, $d_{x^2-y^2}$, and $d_{yz}$. It is also clearly shown in the figure that the flat band along A--R path is dominantly contributed from the $d_{x^2-y^2}$ orbital, which is affected by electronic correlation more prominently near the itinerant-to-localized boundary than the other orbitals, as discussed in the main text and Appendix~\ref{sec:Z}.

\section{Quasiparticle weight $Z$ for different $U$ and $J$ parameters\label{sec:Z}}
\begin{figure}[h!]
    \centering
    \includegraphics[width=0.8\linewidth]{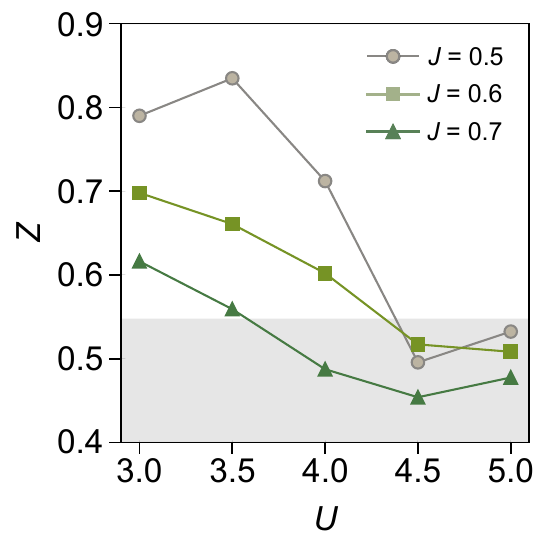}
    \caption{Qausiparticle weight $Z$ of $d_{x^2-y^2}$ orbital obtained from PM DFT+DMFT calculation for pristine \RO. }
    \label{fig:Z}
\end{figure}

Fig.~\ref{fig:Z} shows the evolution of the quasiparticle weight $Z$ of Ru-$d_{x^2-y^2}$ orbital as a function of $U$ for $J=0.5, 0.6$, and  $0.7$ values in line-dot graph. The full data in heatmap with all $U$ and $J$ values calculated in this work is shown in the main text Fig.~\ref{phasediagram}(c). In this figure, the horizontal gray box visualize localized regime at large $U$ or large $J$ limit, discussed in the main text. As discussed in the main text, $Z$ rapidly decreases as $U$ increases at smaller $U$ region but very slowly decreases at larger $U$ region, which is expected from the Hund's correlation effect in multi-orbital system. This saturation of $Z$ is more clearly seen in larger $J$ cases, where $Z$ at smaller $U$ region is also much smaller.


\end{document}